\documentclass{aa}  

\usepackage{graphicx}
\usepackage{txfonts}
\usepackage{color}
\usepackage{wasysym}
\usepackage{natbib}
\newcommand{\aliphCC}{$>\hspace{-0.40cm} - $C$-$C$< \hspace{-0.32cm} - $}

\begin{document}

    \title{Quantum confinement and carbon nanodots:}
        \subtitle{A conceptual view for the origin of diffuse interstellar bands}
    
        \author{A.P. Jones}

     \institute{Universit\'e Paris-Saclay, CNRS,  Institut d'Astrophysique Spatiale, 91405, Orsay, France.\\
               \email{anthony.jones@universite-paris-saclay.fr}              }

    \date{Received 24/01/2025 : accepted 15/05/2025}

   \abstract
{The nature of the Diffuse Interstellar Band (DIB) carriers is perhaps the most studied, longest standing, unresolved problem in astronomy. 
While four bands have been associated with the fullerene cation (C$^+_{60}$) the vast majority ($> 550$) remain unidentified.}
{This works is an attempt to provide a conceptual framework for the typical energy transitions that are central to explaining the origin of DIBs, however, it does not make an association between these transitions and any particular DIBs.}
{The effect of quantum confinement on excitons is used, including charge transfer excitons, to construct a generic basis for the electronic transitions that could, in principle, be coherent with the energies associated with DIBs. In this model the carriers are carbon nanodots (CNDs) modelled as nanodiamonds and a-C(:H) nanoparticles.}
{These preliminary results seem to show that particle size dependent effects in nanodiamond and a-C(:H) CNDs could be consistent with the positions of, and intervals between, some of the DIBs. One particular strength of the model is that predicts  single bands from the majority of single-size particles and, at most, two bands from some of these same carriers. 
In the latter case the two bands come from different transitions and may or may not inter-correlate, depending upon the local environment.}
{This generic framework indicates that the size dependent fundamental transitions in CNDs could provide a viable scenario for the origin of some DIB-type bands. While this work does not identify a single DIB, it furnishes a conceptual view for the DIB origin, and suggests that a more refined exploration of quantum confinement size effects, and exciton physics within the astronomical domain might prove fruitful. This work also hints at the requirement for stable configurations for particular size domains in order to explain DIB wavelength stability.}


   \maketitle

\section{Introduction}
\label{sect_intro}

Other than the inclusion of stochastic heating, a quantum effect due to particles with low heat capacities compared to the energy of an absorbed photon, it appears that quantum processes have been rather neglected in interstellar dust studies.  This situation holds despite the long history of effects that ought to be manifest by dust in the interstellar medium (ISM). Principal among these are the critical effects arising from quantum confinement (QC). This arises through the imposition of dimensional and spatial constraints on photons, excitatiions and atomic constituents at nanometre size scales. Many of the interesting processes relate to the role of excitons, including charge transfer (CT) excitons, in insulating and semiconductor materials under QC effects. 

The diffuse interstellar band (DIB) problem has been around, and unresolved for more than one hundred years. For comprehensive reviews of all of the key DIB related issues the reader is referred to the review by \cite{1995ARA&A..33...19H} and to the proceedings of the ``The Diffuse Interstellar Bands'' IAU meeting No. 297 held at Noordwijkerhout in May 2013 \citep{2014IAUS..297}. These works give an essential state-of-the-art view of DIBs. 

To date, of the more that five hundred and fifty DIBs that have been observed \cite[e.g.,][]{2019ApJ...878..151F}, only four have been unequivocally assigned:  the 936.59, 942.85, 957.75, and 963.27\,nm DIBs are due to C$_{60}^+$ \citep{2015Natur.523..322C,2015ApJ...812L...8W}. Nevertheless, the impressive detective work of associating a single carrier with four DIBs hardly constitutes a satisfying resolution of this outstanding astrophysical problem. 

This paper is an attempt to shed some new light on the DIB identification problem through a conceptual framework involving carbon nanodots modelled as nanodiamonds and as a-C(:H) nanoparticles. It appears that excitonic transitions in these types of particles, critically coupled with QC effects, could be a way explaining the origin of some DIBs. However, this work does not make any specific association between nanoparticle exciton transitions and any given DIBs.

This paper is constructed as follows: 
Section \ref{sect_general} summarises the nature of carbon nanodots,  
Section \ref{sect_excitons} presents the basics of excitons, quantum wells and quantum confinement, 
Section \ref{sect_model} describes a conceptual model for generic DIB-type carriers, 
Section \ref{sect_discussion} discusses the context, and
Section \ref{sect_conclusions} concludes.

\section{Carbon nanodot characteristics}
\label{sect_general}

Interestingly, carbon nanodots (CNDs), sometimes also referred to as carbon dots, were only discovered in 2004, and then quite by chance, by \cite{2004JAmChemSoc...126.12736} during their purification of single-wall carbon nanotubes. Since then, and as highlighted in the review articles by \cite{Giordano_etal_2023}, \cite{Kumar_etal_2022}, \cite{Liu_etal_2020}, \cite{Liu_2020} and \cite{Ozyurt_etal_2023}, CNDs have generated an enormous amount of interest. They are an intriguing class of 0-dimension, quantum-confined systems that are highly stable, chemically inert, simply-synthesised, good conductors, have variable optical properties, and exhibit low toxicity, strong fluorescence,  high photo-stability, and resistance to photobleaching. These properties and, in particular, their strong and tuneable fluorescent emission, generally in the visible to near-infrared wavelength range, enable a wide range of uses in biomedicine, optronics, catalysis, and sensors. These are but a few examples of the myriad applications of the properties of CNDs, which are principally due to size- and structure-dependent excitation effects operating at nanometre scales. 

Unless otherwise specified the following summaries of carbon nanodot properties are gleaned from the selected reviews \citep{Giordano_etal_2023,Kumar_etal_2022,Liu_2020,Liu_etal_2020,Ozyurt_etal_2023} that describe in detail their important properties, fabrication and uses.
The sub-classification of carbon nanodots is complex, ambiguous and far from widely accepted. It seems that a structure-based approach is perhaps the most amenable but even this is imprecise. The following short list attempts to summarise their major classes or types: 
\begin{itemize}
\item GQDs, graphene quantum dots consist of graphene sheets.
\item The following are considered carbon nanodot subgroups: 
\begin{itemize}
\item CQDs, carbon quantum dots, multiple graphene layers, 
\item CNDs/CNPs, carbon nanodots/carbon nanoparticles are spherical, amorphous, containing small aromatics, and   
\item CPDs, carbon polymeric dots, a heterogeneous group.   
\end{itemize}
\end{itemize}
The subclasses of carbon nanodots are usually arrived at through the characterising techniques of Raman spectroscopy in conjunction with infrared and x-ray luminescence.

In what follows, and in the light of the type(s) of carbon nanodots of most relevance to astrophysics, for convenience the CND acronym is used  to describe carbonaceous nanoparticles, unless the reference is to a specific class of carbon nanodot. We do, however, recognise that within the interstellar dust context there is a non-negligible overlap in interest between the CND and CPD classes and this will be implicit in what follows, even though we only refer to them as CNDs. 
 
It is apparent from the literature reviews of CNDs \cite[e.g.,][]{Giordano_etal_2023,Kumar_etal_2022,Liu_2020,Liu_etal_2020,Ozyurt_etal_2023} that, not surprisingly, there is currently a lack experimental and theoretical work that would be of particular relevance to nanoparticles in the ISM. For example, the majority of the studies relate to the emission bands produced by CNDs, mostly at room temperature, whereas in interstellar dust studies we are most interested in absorption from the ground state at cryogenic temperatures ($10-50$K). Thus, the lack of experimental data on the low temperature properties of CNDs and  the fine details of their UV-NIR absorption bands is currently a hinderance to evaluating their relevance to the question of the origin of the majority of DIB-like bands.

\subsection{Structure}
\label{sect_CD_structure}

CNDs generally consist of core/shell structures, with radii in the range $0.5-5$\,nm, corresponding to $\sim 10^2-10^5$ atoms per particle, and with cores sizes of the order of $\sim 1-2$\,nm. The shells are typically amorphous with abundant C$=$C and C$=$O structures, with amino, epoxy/epoxide, hydroxyl, carbonyl, carboxylic acid, epoxy ether functional groups and /or polymer chains. They can be heteroatom doped with O, N, S, P, and F atoms, often incorporated into conjugated $sp^2$ aromatic domains. Particularly important are O and N atoms which can affect their structural and optical properties and promote intramolecular CT.

To highlight the complex overlap in CND properties, as detailed by \cite{Giordano_etal_2023}: CPD structures are much more complex than for other types. 
They exhibit core/shell structures with the cores of dispersed aromatic clusters connected or bridged by $sp^3$ hybridised carbons, while the shells are dominated by chemical functional groups, such as $>$C$=$O, $-$C$-$O$-$C$-$, $-$C$\leqslant_{\rm O}^{\rm OH}$, $-$OH, $-$N$<^{\rm H}_{\rm H}$, and/or polymeric moieties. 
In CPDs it is the core aromatic domains that are the source of fluorescence, with QC effects the main controlling factor in determining their emission. 

\cite{scott_etal_2024} point out the majority of CNDs are complex structures with a mixture of sp$^2$ and sp$^3$ carbons and heteroatoms,  which can be found in mixed crystalline, polycrystalline, polymeric, and amorphous phases within the same nanoparticle. This inherent complexity makes the prediction of their optical properties difficult using simple size-confinement rules but it turns out that their structural geometry affects their emission just as significantly as size \citep[e.g.,][]{scott_etal_2024}. 

The kinds of structures associated with CPDs and CNDs are remarkably similar to that proposed for interstellar carbonaceous nanoparticles  by 
 \cite{2012A&A...540A...1J,2012A&A...540A...2J,2012A&A...542A..98J}, and extended in \cite{2016RSOS....360221J,2016RSOS....360223J,2016RSOS....360224J}, and which are the basis of the carbon nanoparticles in the THEMIS interstellar dust model, The Heterogeneous dust Evolution Model for Interstellar Solids, \citep{2013A&A...558A..62J,2014A&A...565L...9K,2017A&A...602A..46J,2016A&A...588A..43J,2016A&A...588A..44Y,2024A&A...684A..34Y}.

\begin{figure}[t]
\centering
 \includegraphics[width=9.2cm]{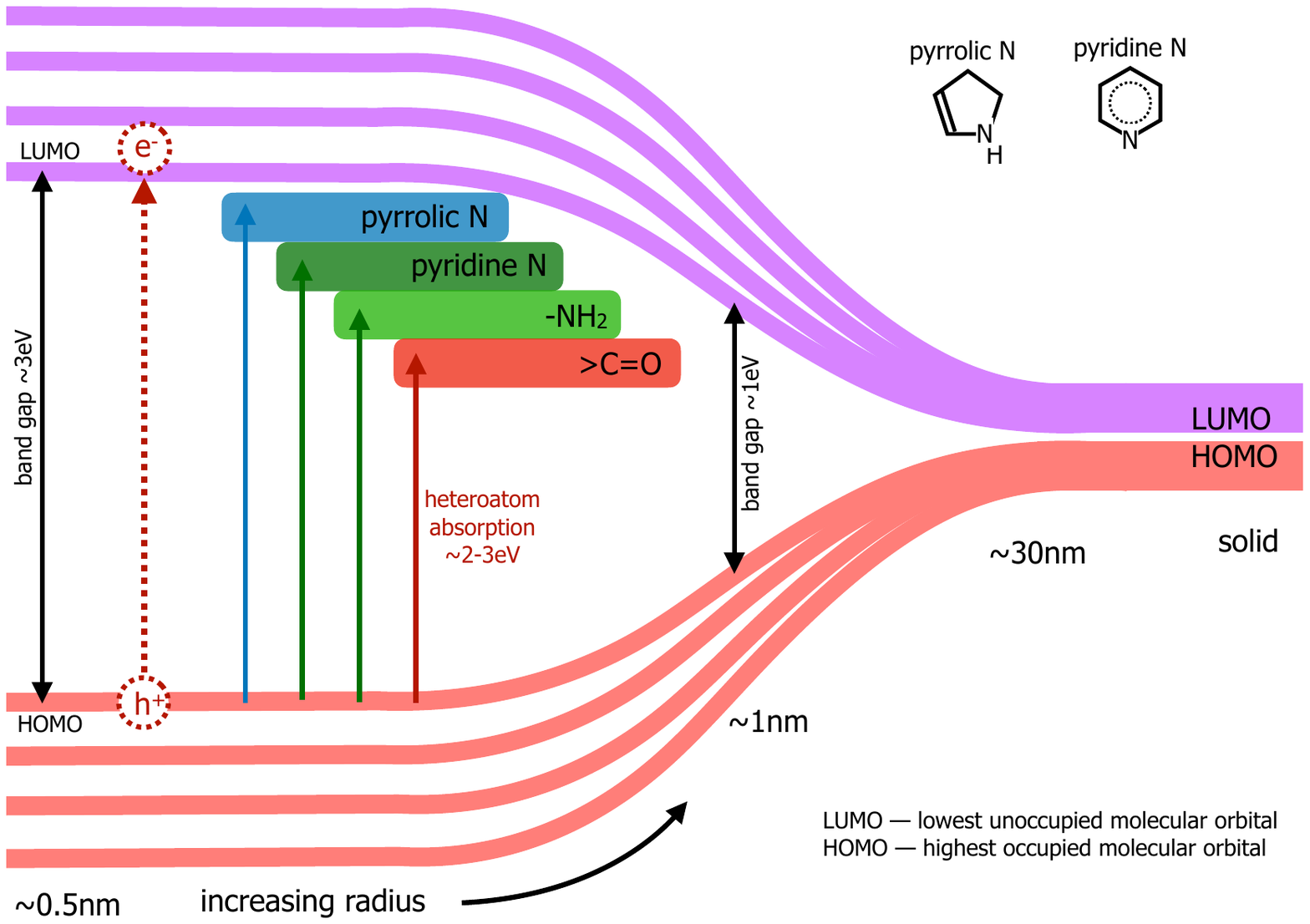}
 \caption{A  schematic of the band structure in semiconducting CNDs, indicating the splitting of the VB and CB into discrete levels with decreasing radius, the size-dependent band gap, exciton ($e^--h^+$) formation, and the introduction of N and O heteroatom-induced energy levels. }
 \label{fig_E_bands}
\end{figure}

\begin{figure}[t]
\centering
 \includegraphics[width=9.0cm]{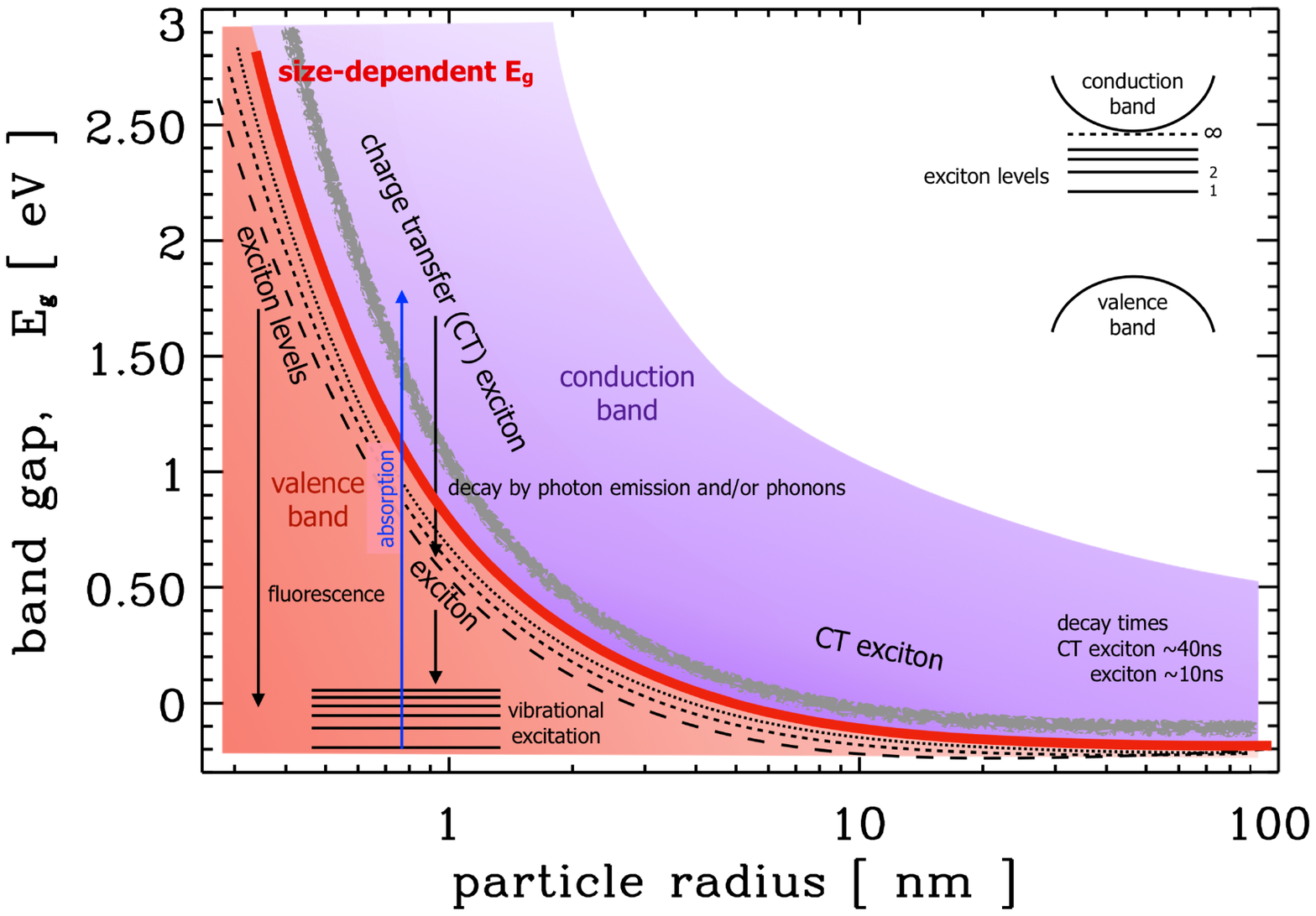}
 \vspace*{-0.3cm}
 \caption{An illustrative view of (charge transfer) exciton transitions and their energies as a function of particle radius. The band gap for a-C(:H) particles as a function of radius, $E_{\rm g}(a) = \{ (1/a[{\rm nm}]) - 0.2 \}$\,eV, is indicated by the red line \citep{2012A&A...542A..98J}.}
 \label{fig_intro}
\end{figure}

\subsection{Quantum effects and optical bands}
\label{sect_VBCB}

In absorption all CND classes have similar optical properties with strong UV absorption extending to the visible and even into the near IR. They show two major peaks: $\sim 200-280$nm due to core $\pi-\pi^\star$ aromatic C$=$C and C$=$N bonds and a second peak at $\sim 300-400$nm due to $n-\pi^\star$ associated with C$=$O. Longer wavelength features ($\lambda > 400$\,nm),  that can overlap with $n-\pi^\star$, are due to surface states containing  lone pairs of electrons.

The valence and conduction bands (VB and CB) in CNDs are broadened and shift from continuous to discrete levels (see Fig. \ref{fig_E_bands}), with respect to solids. The optical gap of CNDs correlates with the extent of the $\pi$-electron systems, as in THEMIS \citep{2012A&A...542A..98J}, and the surface states arising from, principally, heteroatom functionalisation. They show size dependent optical gaps, in part due to QC effects, with the gap increasing as particle size decreases (see Figs. \ref{fig_E_bands} and \ref{fig_intro}). However, the band gap can also be decreased by increasing the O atom content. 

An example of relevant quantum effects are the numerous studies of CNDs and, in particular, nanodiamonds which exhibit clear QC effects. For example, QC in nano\-diamond thin films was demonstrated by \cite{1999PhRvL..82.5377C} using x-ray-absorption spectroscopy. They showed that the particle properties deviate from those of the bulk material for $a < 1.8$\,nm and that the effective mass of exciton electrons $m^\ast = 0.10\pm0.02$\,m$_{\rm e}$, where $m_{\rm e}$ is the mass of an electron. This work also showed how the exciton state and the conduction band edge shift to higher energies with decreasing grain radius. The particle size dependent behaviour of the band gap energy, $E_{\rm g}(a)$, shift for diamond can be empirically fit with the expression 
\begin{equation}
E_{\rm g}(a) = 5.5 + 3.76 \left( \frac{ a }{ {\rm [ \, nm \, ] } } \right)^{-2} \ \ {\rm [ \, eV \, ]}.  
\label{eq_Eg_shift_dia}
\end{equation}

\subsection{Fluorescence}
\label{sect_CD_fluorescencee}

In general the fluorescent emission from CNDs arises from their molecular subdomains, surface states and QC. The associated energy levels and electron cloud distribution of $\pi$ and $\pi^\star$ states in the aromatic domains can interact with $\sigma$ and $\sigma^\star$ states of the $sp^3$ carbon around them. Fluorescent emission can also come from chromophores associated with heteroatom-rich regions (see Fig. \ref{fig_E_bands}), but does not necessarily require the presence of aromatics,  and is aided by the reduction in mobility due to steric hindrance introduced by the covalent bonds. For those interested a wider discussion of the emission properties can be found Appendix \ref{sect_CD_emission}.  

It has been noted that for CQDs with radii increasing from $\sim 1$\,nm to $\sim 3$\,nm the fluorescent emission shifts from the blue and into the red. In another study on CNDs with radii increasing from  $\sim 3$ to 5 to 8\,nm it was found that the optical gap decreases from 3.0 to 2.9 to 2.2\,eV, corresponding to a colour shift from the violet domain into the yellow. Further, it is possible to tune the fluorescent emission from the blue to red wavelength region in multicolour band gap fluorescent CQDs \citep[MCBF-CQDs,][]{MCBF-CQDs}.  Thus, the redshift in emission with increasing particle size appears to be a fundamental property of all classes of carbon nanodots. 

In a study of the role of size, shape, and 3D structure on the emission properties of more than fifty six-ringed benzenoid hydrocarbons \cite{scott_etal_2024} showed that their emission properties can be affected significantly. However, it appears that their absorption properties are much less affected by topology. Given that DIBs are absorption bands, this interesting computational chemistry work on the emission properties of polycyclic aromatic hydrocarbons may help to shed some  light onto the role of the aromatic moieties within hydrocarbon nanoparticles.

\section{Excitons and quantum confinement}
\label{sect_excitons}

Generally, excitations can be free in three dimensions (the soiid state), two dimensions (quantum wells), one dimension (quantum wires) or zero dimensions (quantum dots). An exciton is formed by the absorption of a photon, of energy greater than the band gap ($E_{h \nu} > E_{\rm g}$, see Figs. \ref{fig_E_bands} and \ref{fig_intro}),  that leads to an electron-hole pair ($e^-h^+$) that is bound by an electrostatic Coulomb force. Excitons produce narrow spectral lines in optical absorption, reflection, transmission and luminescence with energies below the free-particle band gap of an insulator or a semiconductor. In the following we briefly summarise the physics of excitons and QC. 

In an exciton the electron resides in the conduction band where its interaction with the hole is screened by the electrons surrounding the hole. It is an electrically neutral quasi-particle in insulators and semiconductors that transports energy but not electrical charge. Where the material dielectric constant is small the $e^- \, h^+$ interaction is strong and the exciton small, as in the case of Frenkel excitons; the opposite holds for Wannier-Mott excitons. In small band gap semiconductors, with generally larger dielectric constants, the electron screening is larger and Wannier-Mott excitons therefore have larger radii and lower binding energies ($r \sim 10$\,nm, $E \sim 0.01$\,eV) than Frenkel excitons ($r \gtrsim 0.1$\,nm, $E \sim 0.1-1$\,eV), the latter can be bound to a single atom. Nevertheless, it is possible for different types of excitons to exist in the same material. The effective mass of the electrons is smaller in semiconductors, as for amorphous (hydro)carbon materials as opposed to diamond, which is an insulator. 

In a 3D semiconductor bulk material the Wannier-Mott exciton energy, $E_n$ and radius, $r_n$, are given by
\begin{equation}
E_n = - \left( \frac{ \mu }{ m_{e,0} \, \epsilon^2 } {\rm Ry}  \right) \left( \frac{ 1 }{ n^2 } \right)  
\ \ \ \ \ \ \  {\rm and} \ \ \ \ \ \ \ \ 
r_n = \epsilon \, \left( \frac{ m_{e,0} }{ \mu } \right) \, a_0  \, n^2,   
\end{equation}
where $m_{e,0}$ is the electron mass, $\epsilon$ is the dielectric constant\footnote{Typical values for a dielectric constant (static relative permitivity) at room temperature  are: 2.1 for PTFE/Teflon, 2.25 for polyethylene, $5.5-10$ for diamond, and $10-15$ for graphite. For amorphous carbons the relative permitivity is likely in the range $4-8$ \citep{2012A&A...540A...2J}.}, $n$ is the principal quantum number, and $\mu$ is the reduced mass, $\mu^{-1} = m_e^{-1} + m_h^{-1}$ with $m_e$, and $m_h$ the effective masses of the electron and hole, respectively.  The Rydberg unit of energy, ${\rm Ry}$, and Bohr radius, $a_0$ , are 
\begin{equation}
1 \, {\rm Ry} = \frac{ m_{e,0} \, e^4 }{ 8 \, \epsilon_0^2 \, h^2 } = 13.6057 \, {\rm eV}, 
\end{equation}
\begin{equation}
a_0 = \frac{ 4 \pi \epsilon_0 \, \hbar^2 }{ m_{e,0} \, e^2 } = 5.2918 \times 10^{-9} \, {\rm cm} = 0.052918 \, {\rm nm},  
\end{equation}
where $\epsilon_0 \ (= 8.8542 \times 10^{-12}$\,F\,m$^{-1})$ is the permittivity of free space.  
As an illustration, in a-C(:H) materials, with $\epsilon \simeq 6$ and assuming $m_e = m_h = 0.07\,m_{e,0}$, this equates to an exciton ground state energy of $0.1$\,eV and radius of $11$\,nm. For diamond (graphite), assuming $\epsilon \simeq 8.5$ (12.5), and $m_e = m_h = 0.1\,m_{e,0}$, this equates to an exciton ground state energy of $0.01$\,eV ($0.004$\,eV) and radius of $8.5$\,nm ($13$\,nm). These numbers, relating to bulk materials, indicate the comparatively large radii of Wannier-Mott excitons, which are larger than or comparable to typical interstellar nanoparticle radii ($0.4-10$\,nm). Table \ref{table_parameters} summarises the key material parameters; showing the values for graphite (a conductor), diamond (an insulator) and a-C(:H) (a semiconductor).

\begin{table*}
\caption{Adopted nanoparticle model and exciton parameters and the calculated values.}
\centering
\begin{tabular}{lccccccc}
      &         &        &      &    \\[-0.35cm]
\hline
\hline
      &         &        &       &   \\[-0.35cm]
  structure              &    $\epsilon$ [assumed value]   &  $m_w^\star/m_{\rm e,0}$     & $E_{\rm ex,0}$ [eV] & $r_{\rm ex}$ [nm] &    $\delta a_d$ [nm]   &        $E_{\rm g}$ [eV]         \\[0.05cm]
\hline
                              &                           &              &    &   &                                   &                            \\[-0.30cm]
 nanodiamond       &     5.5--10 \ \ [8.5]    &     0.1   &   0.01  &  8.5 &             0.044                &     $5.5+(3.76 / a^2)$  \\
a-C(:H)                  &      4--8  \ \ \ \ \ \  [6]  &     0.07   &  0.1   &  11 &        0.062--0.018          &      $(1/a) - 0.2$  \\
 graphite                &       12.5                   &     0.1   &   0.004  &  13 &             0.241                &     0  \\
\hline
\hline
     &      &         &       &    \\[-0.25cm]
\end{tabular}
\tablefoot{
$\epsilon$ is the dielectric constant. $E_{\rm ex,0}$ and $r_{\rm ex}$ are the Wannier-Mott exciton ground state energy and radius, respectively. $\delta a_d$ is the particle radius discretisation, which is size-dependent for the a-C(:H), shown for radii of $0.4 - 4$\,nm. In the Band gap, $E_{\rm g}$, expressions the radius, $a$, is in nm.
}
\label{table_parameters}
\end{table*}

In bulk material excitons the Coulomb electrostatic potential between the $e^-$ and $h^+$ is $U(d) = -e^2/ (\epsilon \, d)$ (CGS units), where $d$ is the $e^- \, h^+$ separation. The energy levels can be expressed as a modified Rydberg equation because the problem can be considered in terms of a hydrogen-like atom, or more correctly a positronium particle, that is
\begin{equation}
E_n = E_{\rm g} - \frac{ \mu \, e^4 }{ 2 \, \hbar^2 \, \epsilon^2 n^2 },
\label{eq_positronium}
\end{equation}
 where the energy levels are with reference to the top of the valence band \citep{Kittel_1996}. The exciton ground state energy ($n=1$ in Eq. \ref{eq_positronium}) is also the exciton ionisation energy (see the upper right inset in Fig. \ref{fig_intro}).  
At low temperatures, the higher excitonic levels approach the band edge \citep{2014Natur.514..343K} and form a series of  absorption lines similar to those of hydrogen. 

The fact that weakly bound Wannier-Mott excitons in the bulk state can have dimensions of the order of 10\,nm or more obviously imposes critical constraints on excitons in nanoparticles ($a \simeq 1$\,nm). Excitons can in fact be located on the same molecule, as occurs in fullerenes, and the typical exciton binding energy in molecular crystals composed of aromatic molecules, such as anthracene or tetracene, is $0.1-1$\,eV. CT excitons, where the electron and hole are on neighbouring moieties, which are typically aromatic, are typical of organic materials (see Appendix \ref{sect_CT_DIBs} for a brief discussion). It is also possible for molecular excitons, and CT excitons to interfere.

\subsection{Quantum confinement (QC)}
\label{sect_confinement}

Turning to the case of excitations in solids that are bound in three dimensions and the effects of QC.\footnote{The energy level system of quantum wells, where the particle is confined in two dimensions, is briefly outlined in Appendix \ref{sect_wells} but is not considered here because it is not a suitable approach for nanoparticles. However, this case could perhaps be applied to thin mantles on 3D particles or to 2D particles such as graphite platelets (graphene sheets).} This is the squeezing of excitons in semiconductor nanoparticles, with radii smaller than their exciton Bohr radius, which means that the energy levels can be predicted using the ``particle in a box'' model where, in this case the states will depend upon the constraining linear dimensions or diameter of the particle (a nanograin), that is the size of the box. The band gap. $E_{\rm g}$, decreases as the degree of confinement increases, that is with decreasing size, because the energy levels split. However, for a-C(:H) semiconductor nanoparticles the band gap actually increases with decreasing size due to aromatic domain size limitations \citep{2012A&A...542A..98J}, which means that the latter effect, an increase in $E_{\rm g}$, must actually be somewhat offset by the former effect (energy level splitting). This is schematically illustrated in Figs. \ref{fig_E_bands} and \ref{fig_intro}. 

In order to study the size dependence of exciton energetics the $e^- \, h^+$ pair of the exciton can be thought of as a hydrogen atom in the Bohr model, with the nucleus replaced by a hole, the solution to the particle in a box for the ground level ($n = 1$), using the reduced mass, gives the energy levels of the exciton. The Bohr radius of an exciton is given by 
\begin{equation}
a_0^\ast = \epsilon_r \left( \frac{ m }{ \mu } \right) a_0
\end{equation}
where $m$ is the carrier mass, $\mu$ is the reduced mass and $\epsilon_r$ is the linear dimension(radius)-dependent dielectric constant (relative permittivity).

The exciton binding energy results from the $e^- \, h^+$ Coulomb attraction: the negative energy of attraction is proportional to Rydberg's energy and inversely proportional to the square of the size-dependent dielectric constant of the semiconductor. The radius of a quantum dot determines the confinement energy of the exciton and when it is smaller than the exciton Bohr radius, the Coulomb interaction must be modified.

The band gap energy, $E_{\rm g}$, confinement energy, $E_{\rm conf}$, and exciton binding energy, $E_{\rm ex}$, energies are related by
\begin{equation}
E_{\rm conf} = \frac{ \hbar^2 \pi^2 }{ 2 \, \mu \, a^2 } 
\end{equation}
\begin{equation}
E_{\rm ex} = - \frac{ 1 }{ \epsilon^2 } \frac{ \mu }{ m_e } {\rm Ry} = - {\rm Ry^\ast}
\end{equation}
\begin{equation}
E =  E_{\rm g} + E_{\rm conf} + E_{\rm ex} = E_{\rm g} + \frac{ \hbar^2 \pi^2 }{ 2 \, \mu \, a^2 } - {\rm Ry^\ast}
\end{equation}
where $E$ is the energy of the bound exciton, $a$ is the radius,\footnote{Given that particle dimensionality (1D, 2D or 3D) is critical, the `radius' quoted here is probably no more than a proxy for the largest particle dimension. Hence, the number of atoms per particle may be less than that for the idealised spherical CNDs considered later.} $\mu = m_e m_h /(m_e + m_h)$ is the reduced mass, $m_e$ is the free electron mass, $m_h$ is the hole mass, and $\epsilon_r$ is the size-dependent dielectric constant. These simplified equations imply that the electronic transitions of quantum dots depend on radius and underline the fact  that QC effects are only apparent below some critical radius ($\simeq 1-2$\,nm) and never in larger particles. 

Considering the energetic properties of CNDs in more detail, the wave vector of a given state is defined by 
\begin{equation}
k_n = \left( \frac{ n \, \pi }{d} \right)^2, 
\label{eq_kn}
\end{equation}
where $n \geq 1$ is the integer quantum number, $d$ is the width of the quantum well and the exciton energy levels are  given by 
\begin{equation}
E_n = \frac{\hbar^2\,k_n^2}{2\, m_w^\ast} = \frac{\hbar^2 }{2\, m_w^\ast} \left( \frac{ n \, \pi }{d} \right)^2, 
\end{equation}
where $\hbar = h/2 \pi$ with $h$ the Planck's constant and $m_w^\ast$  is the effective mass of the carriers within the well region.
The transition energy from the $n^{\rm th}$ level to the ground state ($n=1$) is therefore 
\begin{equation}
\delta E_n = \frac{1}{2} \frac{\hbar^2 }{m_w^\star} \left( \frac{ \pi }{d} \right)^2 \ \left(  n^2 -1 \right), 
\label{eq_deltaE}
\end{equation}
corresponding to a wavelength  
\begin{equation}
\lambda_n  = \frac{2 \, c \ m_w^\star \, d^2}{h \,(  n^2 -1 )} = \frac{8 \, c}{h} \ \frac{ 1 }{(  n^2 -1 )} \ \{ m_w^\star \, d^2 \},  
\label{eq_product}  
\end{equation}
where the bracketed term is the product for the case where the energy or wavelength of a transition is well determined, as in the case of the DIBs, but neither the particle size nor the effective mass of the excitation carrier are known. For the fundamental ($n=2 \rightarrow n=1$) transition the energy and wavelength are 
\begin{equation}
\delta E_n = \frac{3}{2} \frac{\hbar^2 }{m_w^\star} \left( \frac{ \pi }{d} \right)^2 
\hspace{0.7cm} {\rm and} \hspace{0.7cm} 
\lambda_n  = \frac{8}{3} \frac{c}{h} \ \{ m_w^\star \, d^2 \}.    
\end{equation}

\begin{figure} 
\centering
 \includegraphics[width=9.5cm]{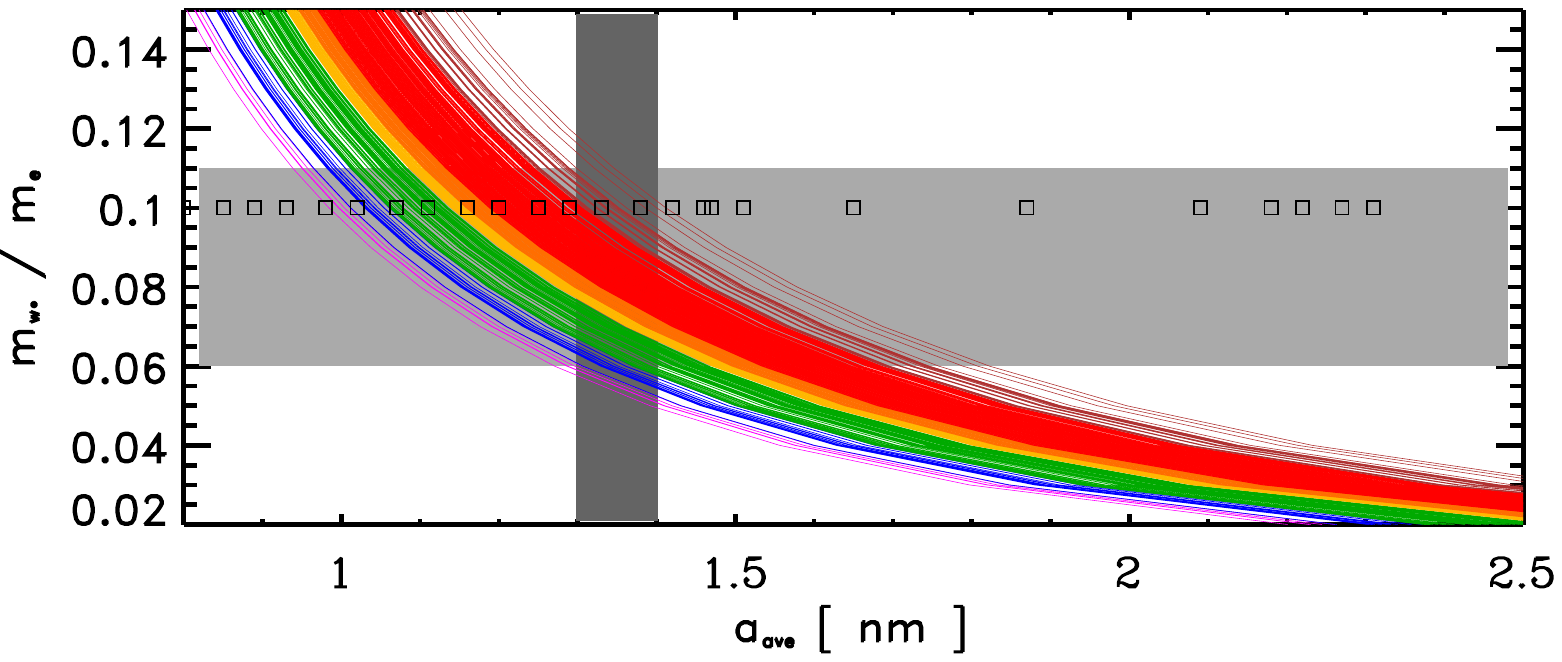}
 \vspace*{-3.2cm}
 \caption{The product $\{ m_w^\star \, d^2 \}$, derived from Eq.\,(\ref{eq_product}), for over 550 DIB wavelengths \citep{2019ApJ...878..151F}. The horizontal grey band shows typical values for the effective carrier masses ($m_w^\star = 0.06-0.11$). The squares indicate the radii of the diamond network particles \citep[][see Sect. \ref{sect_model}]{2020_Jones_nd_CHn_ratios}  for $m_w^\star = 0.1\,m_e$.The vertical dark grey band marks the mean radius range for pre-solar nanodiamonds.}
 \label{fig_CND_1}
\end{figure}

The product $\{ m_w^\star \, d^2 \}$ is shown in Fig. \ref{fig_CND_1} for $\sim 550$ observed DIB wavelengths \citep{2019ApJ...878..151F} , assuming that all of the transitions from the $n=2$ to 5 levels end in the ground state ($n=1$). This figure shows that for the typical range of carrier effective masses ($m_w^\star = 0.06-0.11$) the purple to infrared DIBs (represented by the violet to brown colours) would, for example, be compatible with nanodiamond radii $\sim 0.9$\,nm to $\sim 1.8$\,nm \citep{2020_Jones_nd_CHn_ratios} from the shortest to longest wavelength DIBs, respectively, corresponding to particles containing $\sim 500$ to several thousands of C atoms. This example illustrates the phenomenological importance of CNDs and is  not meant to imply that nanodiamonds are at the origin of the DIBs.

\subsection{Quantum Dots}
\label{sect_CNDs}

Quantum dot (QD) properties are intermediate between bulk semiconductors  and atoms or molecules. They are semiconductor nanoparticles, with radii of the order of a few nm, which exhibit optical and electronic properties that differ from larger particles because of quantum effects. QDs absorb UV photons leading to electron excitation from the valence band (VB) to the conduction band (CB),  the transition back to the VB results in photoluminescence the energy of which depends on the CB to VB gap or, where band structure is no longer a good description, to discretised energy states.  As particle size decreases the VB and CB band of solids divide into discrete levels. 
Electron excitation promotes an electron to the CB, leaving a hole in the VB, forming an an exciton (a bound $e^- \, h^+$ pair). Electron-hole recombination leads to fluorescent emission as shown schematically in Fig. \ref{fig_intro}). However, as Fig. \ref{fig_E_bands} illustrates, there are other energy levels within the band gap that can also play a role. 

The emitted photon energy is the sum of the band gap energy between the highest occupied level (HOMO) and the lowest unoccupied energy level (LUMO) (see Fig. \ref{fig_E_bands}), the confinement energies of the hole and electron, and the exciton binding energy, 
\begin{equation}
E_{h \nu} = E_{\rm g} + E_{\rm conf} + E_{\rm binding}.  
\end{equation}
In general QDs with radii of the order of a few nm emit in the orange or red regime and smaller QDs ($a \sim 1$\,nm) emit blue and green light (e.g. see Fig. \ref{fig_CND_1}) with the colour depending on the particular composition of the QD. Such particles have  applications in quantum computers, solar cells, LEDs, lasers, and medical imagers, to name but a few of their myriad uses. Thus, it is to be expected that CND emission, as per QD emission, is rather insensitive to the exact nature of the material but is  dominated by size effects (QC). It is, however, clear that structural effects can also play a critical role \citep[e.g.,][]{scott_etal_2024}.

\subsection{Indiications for interstellar dust studies}
\label{sect_IS_dust}

\begin{figure} 
\centering
 \includegraphics[width=9.2cm]{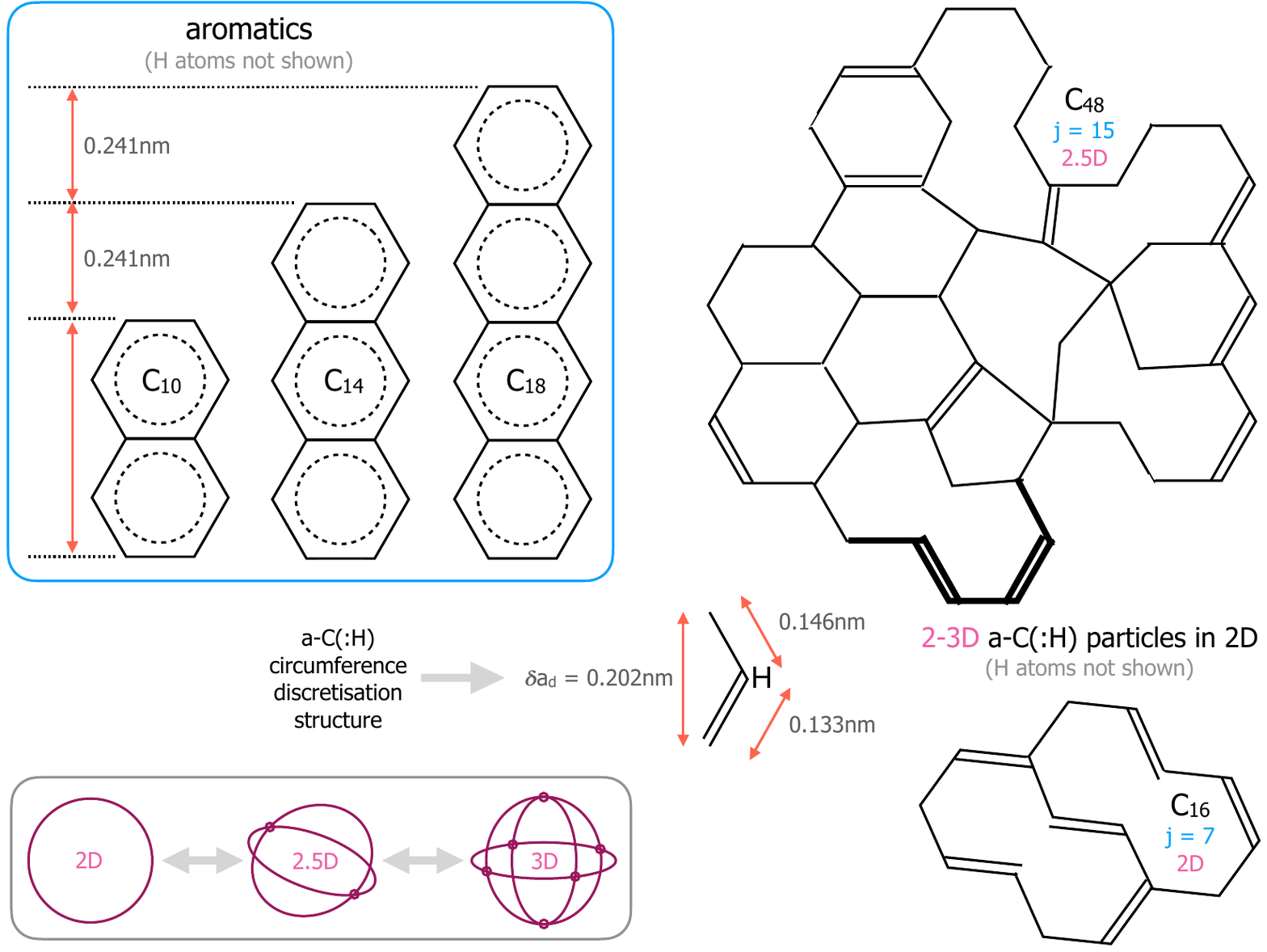}
 \caption{A schematic view of the essential compositional and structural elements in the CND modelling (above and right), along with the key bond lengths and particle dimensionality (bottom left, see Appendix \ref{sect_aCH_CND}).}
 \label{fig_structures}
\end{figure}

Given that much interstellar dust is in the nanoparticle size regime it seems to be an inescapable conclusion that interstellar nanoparticles must exhibit QC effects. This being the case, it seems entirely reasonable to assume that a significant fraction of the DIBs could arise from exciton transitions affected by QC effects. Given that the smaller interstellar nanoparticles only contain several tens to hundreds of atoms, their sizes will be atomically quantised, that is the particular sizes will be determined by the specific chemical composition, structure, and the associated bond lengths. Clearly the structural quantisation level splitting will be more pronounced for the smaller particles and will decrease with increasing size.  

A key issue for astrophysics, and critical to DIBs, is the full width at half maximum (FWHM) of the associated phenomena. The width of the CND fluorescence bands, little discussed in the literature, tend to be broad and independent of the colour. As mentioned above in Section \ref{sect_excitons}, excitons in insulators and semiconductors intrinsically give rise to narrow spectral lines in optical absorption, reflection, transmission and luminescence with widths seemingly of the order of $3-20$\,nm. However, laboratory-measured FWHMs lie in the range  $100-150$nm \cite[e.g.,][]{Kumar_etal_2022,Ozyurt_etal_2023} and are possibly as narrow as several tens of nm \cite[e.g.,][]{Liu_etal_2020}. For example, \cite{Liu_etal_2020_FWHM} synthesised CPDs emitting  a deep red band with an unprecedented FWHM of 20nm and \cite{Yuan_etal_2019} made deep-blue light-emitting CNDs with FWHM of $\sim 35$nm. These FWHMs for emission bands are are still about an order of magnitude wider than the median DIB absorption FWHM of $\simeq 3$nm. 

The measured width of the fluorescence from CNDs arises from the range of exciton levels within a particle, see inset in Fig. \ref{fig_intro}. The energies of the particular levels also depend on structural variations within a given particle, and across a collection of particles of differing sizes (as in the laboratory measurements), which leads to a dispersion in the local dielectric environments and hence to emission bands that are broader than the intrinsic single transition exciton band widths. 
In the case of individual particles, it is logical to assume that the dispersion or the number of local structural/dielectric environments decreases as the particle size decreases leading to fewer available excitation states. Thus, we might expect fewer transitions from smaller particles. However, the energies of the levels will depend upon the particle size, composition, and structure and vary between particles, even for fixed size. For sub-nm particles in the ISM, which are completely isolated, we are also dealing with a collection of particles along any given line of sight, and will observe particle variations even for fixed radius. Thus, for single photon excitation from a ground state, at cryogenic temperatures ($10-50$K), we might expect narrow absorption bands.

\section{A conceptual model}
\label{sect_model}

In ISM studies it is always implicitly assumed that the dust size distribution is continuous. This cannot, however, be the case because its structures are atomic networks with bond lengths fixed by the elements involved and their particular bonding configurations. The following applies this within the framework of carbonaceous particles of a regular diamond lattice and also of amorphous a-C(:H). In order to make comparisons with the experimental work on CNDs ($a \sim 0.5 - 5$\,nm), the properties of particle radii from 0.4 to 4\,nm will be considered. The lower radius limit corresponds to the minimum a-C(:H) dust size in the THEMIS diffuse ISM caset \citep{2013A&A...558A..62J}, Further, the model predicts that 4\,nm radius grains are about an order of magnitude less abundant than 0.4\,nm radius grains. Thus, and within the THEMIS framework, there is a bias towards the smallest grains for any process yielding DIB-like bands. 

\begin{figure} 
\vspace*{-0.5cm}
\centering
 \includegraphics[width=9.5cm]{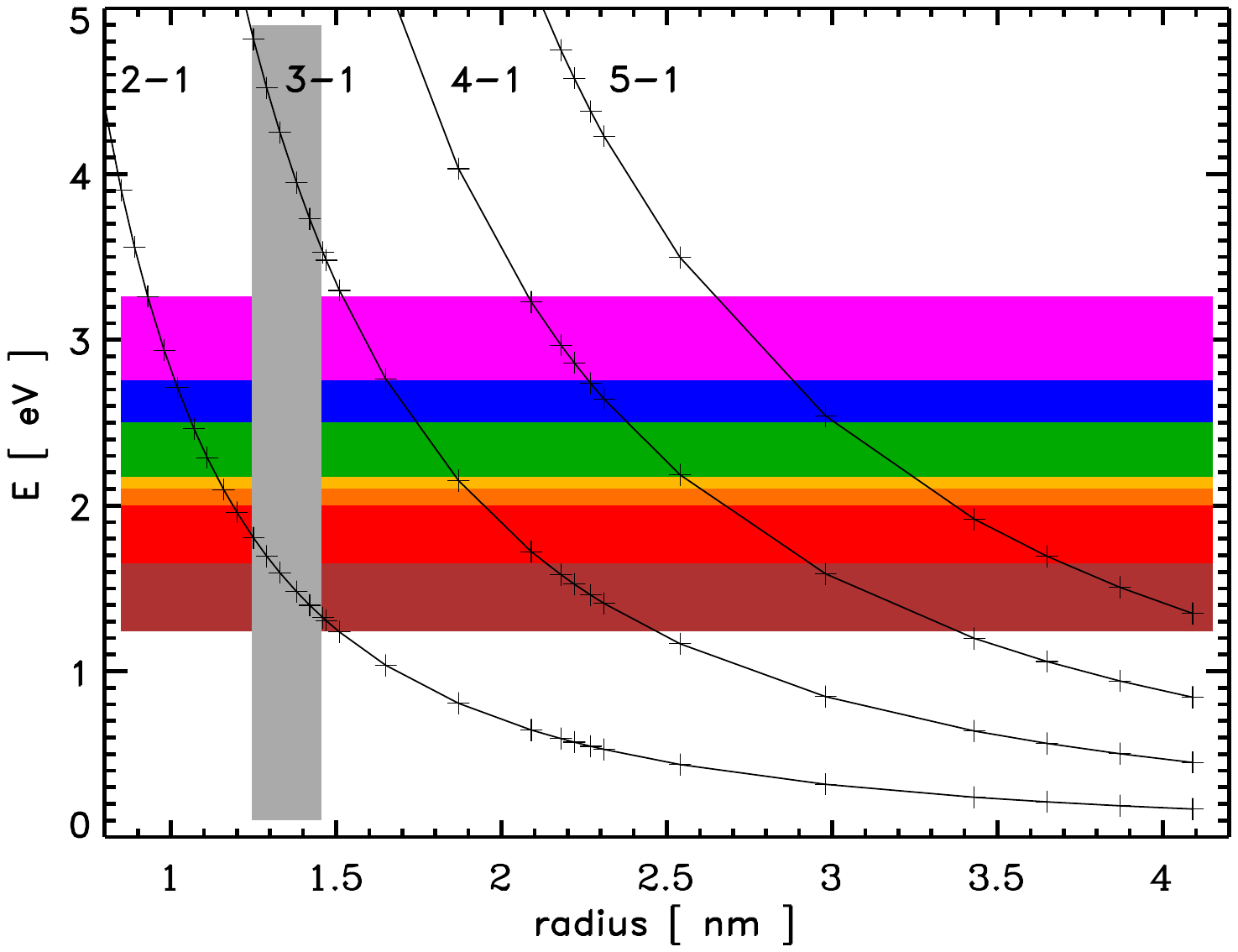}
 \caption{Transition energies for nanodiamond transitions ($\delta a = 0.044$) between levels $n=2-5$, and the ground state ($n=1$) as a function of size. The crosses mark the positions of the modelled nanodiamond radii. The horizontal coloured bands indicate the energies and visible wavelengths of the transitions. The vertical grey band marks the mean radius range for pre-solar nanodiamonds.}
 \label{fig_CND_2}
\end{figure}

The preliminary and conceptual nanodiamond CND model mentioned in Sect. \ref{sect_confinement} was constructed using the theoretically and experimentally determined properties of diamond, and nanodiamonds  \citep{2020_Jones_nd_CHn_ratios,2020_Jones_nd_ns_and_ks}. The contiguous network modelling of diamond nanoparticles elucidated a dimensional periodicity of the order of 0.04 to 0.05\,nm, with an average step value of $\simeq 0.044$\,nm in radii for particles with radii $\lesssim 1.5$\,nm \citep{2020_Jones_nd_CHn_ratios}. This periodicity or discretisation is due to the fact that the diamond network consists of two interlocking face centred cubic (fcc) lattices offset by one quarter of the cubic unit cell edge length (0.357\,nm), which is also the side length of each of the fcc lattices. Thus, a contiguous diamond network can only be built up in diametric increments that ensure `closed' structures, that is with no dangling carbon atom (C$-$) bonds, and this unit is the offset between the two fcc lattices or one quarter of the unit cell dimension (0.089\,nm) equivalent to 0.0445\,nm in radius increments. This is the origin of the periodicity found in the quasi spherical nanodiamond structures generated by  \cite{2020_Jones_nd_CHn_ratios}. This periodicity is somewhat evident  in Fig. \ref{fig_CND_2} where the idealised nanodiamond transition energies, calculated using Eqns. (\ref{eq_kn}) to (\ref{eq_deltaE}),  are shown as a function of particle size, in relation to the visible spectrum wavelengths, assuming a carrier mass $m_w^\star = 0.1\,m_{\rm e,0}$. 

Supposing that, rather than nanodiamonds (Fig. \ref{fig_CND_1}), the significantly more abundant a-C(:H) nanoparticles \cite[e.g.][]{2020_Jones_nd_ns_and_ks} are the carriers of DIB-like transitions, we can focus on their intrinsic aromatic moiety substructures (see Appendix \ref{sect_CT_DIBs}). If the quantum confinement of the excitons in these materials is driven by the aromatic domain size, with the number of aromatic rings per moiety $N_{\rm R} = 1, 2, 3,$\ \dots, then the carrier diameters are quantised by the aromatic ring dimension across the flat edges, $d = ( 0.139 \, \surd3 \, m) = ( 0.241 \, m)$\,nm, where $m$ is the diametric quantisation ($m = 1, 2, 3$ for sub-nm particles) and the aromatic C$-$C bond lengths is taken to be 0.139\,nm. Thus, $d = 0.241, 0.482,$, and 0.722\,nm for $m = 1, 2, 3$, respectively (see top left panel in Fig. \ref{fig_structures}). This molecular structure 2D quantisation step of 0.241\,nm for aromatic moieties can be compared with that of 0.044\,nm for 3D quasi spherical nanodiamonds \citep[][see following section]{2020_Jones_nd_CHn_ratios}. This approach leads to a coarse size quantisation, yields few bands, and is too simplistic for a-C(:H) nanoparticles. Nevertheless, the shape of these aromatic domains does have an important effect on CND emission behaviour but less so on the absorption properties \citep{scott_etal_2024}.

CND phenomena are generic and will therefore also be exhibited by a-C(:H) nanoparticles. Structurally, CNDs consist of small aromatic domains linked by nonconjugated bonds that effectively isolate these sp$^2$ regions and can lead to a bending or twisting of these aromatic moieties, which tends to produce a red-shifting of the bands with respect to expectations \citep{scott_etal_2024}. However, these same authors found that molecular symmetry and peri- or cata-condensation, that is the degree of compactness, were not found to be important factors. As \cite{scott_etal_2024} point out, changes in geometry through the steric twisting of planar molecules into 3D structures significantly changes their emission behaviour and, for fixed size, aromatic molecules can span the entire visible spectrum. 

In the case of a-C(:H) nanoparticles, and because of their shell like structures encompassing aromatic domains \cite[e.g. see][]{2012A&A...542A..98J}, the particle size quantisation is determined by the particle circumference, which can be approximated as an integer number of CC bond lengths of the order of $d_{\rm CC} = 0.139 - 0.154$\,nm where the range is due to the spread of 
aromatic, olefinic, and aliphatic\footnote{Single aliphatic bonds, \aliphCC, form between $sp^3$ hybridised C atoms and double olefinic bonds, $>$C=C$<$, between $sp^2$ hybridised atoms. Aromatic $sp^2$ $\simeq$C$\simeq$C$\simeq$ bonds are intermediate in properties.} bond lengths. It has been shown that the principal colour-determining factor (in emission) is the longest linear dimension of the conjoined aromatic rings within a given system, that is, the maximum length of the acene-like structures \citep[e.g.,][]{scott_etal_2024} such as naphthalene anthracene, tetracene, and pentacene with 2, 3, 4 and 5 linearly-conjoined benzenoid rings (e.g., see upper left panel in Fig. \ref{fig_structures}). In distorted chromophores it is this longest linear acene-like segment that acts as the dominant chromophore \citep{scott_etal_2024}.

Considering a-C(:H) nanoparticles as per the THEMIS model, with aromatic moieties, as mentioned above, but also with sub-structures consisting of conjugated bonding systems, such as C--C=C with bond lengths of 0.146\,nm, and 0.133\,nm, respectively, and a bond angle of  121$^\circ$. The distance between the culminating carbon atoms in this conjugated system is 0.202\,nm (see inset in Fig. \ref{fig_structures}). The following modelling assumes that the a-C(:H) nanoparticle circumference is quantised by twice this distance (0.404\,nm), that is by butadiene units, --C=C-C=C--,  that form peripheral (ring) structures. This is illustrated by the highlighted bonds in the larger a-C(:H) structure  in the upper right of Fig. \ref{fig_structures}.  Appendix \ref{sect_aCH_CND} details how the a-C(:H) nanoparticle circumference discretisation (and hence the radius discretisation) is determined. In this case, and unlike for the nanodiamonds, the discretisation step is size-dependent. The key nanoparticle parameters are summarised in Table \ref{table_parameters}.

\begin{figure*} 
\vspace*{-1.2cm}
\centering 
 \includegraphics[width=19.0cm]{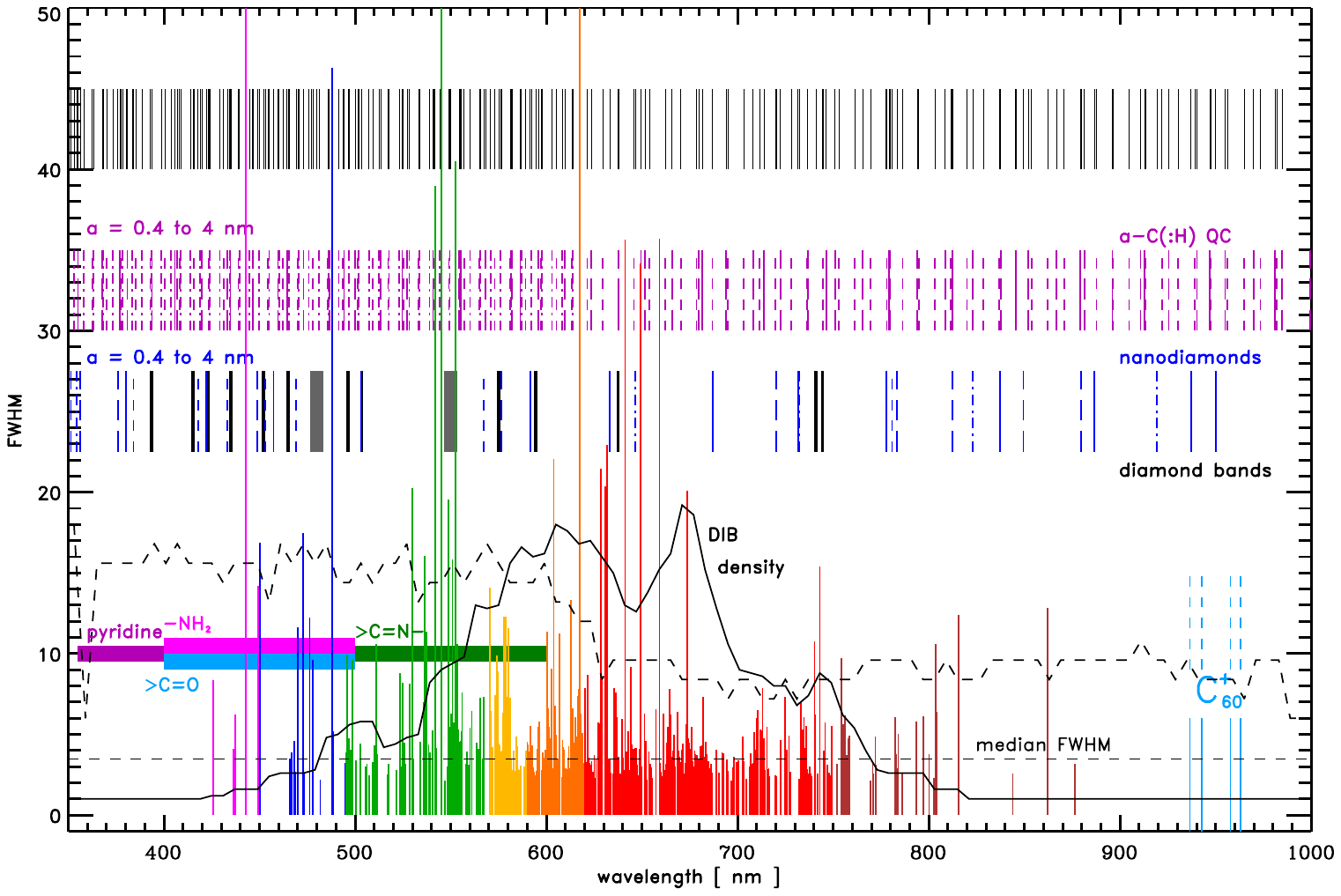}
 \vspace{-2.0cm} 
 \caption{A set of over 550 DIB full widths at half maximum (FWHM) and colours (coloured lines) as a function of wavelength  taken from \cite{2019ApJ...878..151F}. The cobalt blue lines indicate the C$_{60}^+$ DIBs \citep{2015Natur.523..322C,2015ApJ...812L...8W}. The DIB density line schematically shows their number per unit wavelength interval and the dashed black line shows the same for the a-C(:H) particles. The short vertical blue lines indicate the QC transition bands for nanodiamond structures of \cite{2020_Jones_nd_CHn_ratios} 
for transitions from the ground state to $n = 2$ to 5, solid, long-dashed, short-dashed, and dashed-dotted lines, respectively. Also shown in black are the positions of the bulk diamond optical impurity and defect related bands, the two wider dark grey bands indicate the broader bands at 480 and 550nm \cite[see Table 3 in][and references therein]{2020_Jones_nd_ns_and_ks}. The purple lines show the idealised a-C(:H) nanoparticle QC transition bands with the same line transition style key as for the nanodiamond bands. All modelled bands were calculated using Eqns. (\ref{eq_kn}) to (\ref{eq_product}). The topmost thin black lines show, undifferentiated, the modelled and measured bands. The coloured, horizontal, wide bands indicate the CND emission bands for the indicated surface, O and N heteroatom species.}
 \label{fig_CND_3}
\end{figure*}

Fig. \ref{fig_CND_3} shows the positions and full width half maxima for a large number of DIBs taken from the Apache Point Observatory Catalog of 559 Optical Diffuse Interstellar Bands \cite[$4429.00-8763.72\AA$,][]{2019ApJ...878..151F}\footnote{VizieR On-line Data Catalog: J/ApJ/878/151.}  and the work of \cite{2008ApJ...680.1256H,2009ApJ...705...32H}. Superimposed on these are the predicted exciton band transitions ($n = 1$ to $n = 2-5$) for nanodiamonds (blue) and a-C(:H) nanoparticles (purple), calculated using Eqns. (\ref{eq_kn}) to (\ref{eq_product}), and assuming effective carrier masses of 0.1, and 0.07, respectively.  Also shown are the bulk diamond bands (black and dark grey).  The broad bulk diamond impurity/defect bands are indicated in darker grey. In this figure the summed and undifferentiated (nano)diamond and a-C(:H) bands are also shown by the upper short black lines in order to provide an overall picture of the band distribution. The a-C(:H) bands are more closely spaced (purple lines) than for nanodiamond (blue lines)  because of the finer circumference, rather than radius-determined, discretisation steps. Hence, there are many a-C(:H) bands apparent in the wavelength regions populated by the DIBs, especially at the shorter wavelengths. The nanodiamond and a-C(:H) band characteristics are summarised in Table \ref{table_DIB_list}. What this table shows is that the number of nanodiamond and a-C(:H) bands within the wavelength range of the observed DIBs ($\lambda_{\rm DIB} = 425.90 - 876.37$\,nm)  is less that the number of modelled nanoparticle sizes. That the nanodiamond modelling yields fewer band is due to the adoption of the structure calculations of  \cite{2020_Jones_nd_CHn_ratios}, the particle radii for which are shown by the squares in Fig. \ref{fig_CND_1}. From the several overlaps in the radius ranges in Table \ref{table_DIB_list} it is evident that some particles exhibit more than one band. These arise from different transitions but in each case any given particle shows only two bands. Of the 43 modelled nanodiamonds 3 (14\%) show two DIB-like bands, 51\% show no bands, and for the 133 modelled a-C(:H) nanoparticles 28 (26\%) show two DIB-like bands and 18\% show no bands in the DIB wavelength range. The reason that some modelled particles ($a \simeq 0.4$ to 4) show no relevant bands is not that they do not exhibit bands but that those bands lie outside the considered wavelength range ($\lambda \simeq 425.9 - 876.4$\,nm). The indication that some particles exhibit two bands implies that there ought to be some pair-wise correlations in any observed DIB-like bands. However, given that the bands come from different energy transitions, the band pairs may not correlate because of differences in the local ISRF. That most of the modelled particles show only single bands over the DIB wavelength range would appear to be consistent with the observational indication that they arise from unique carriers.

Obviously, the perfect periodicities adopted here arise entirely from the major assumptions about the atomic regularity of the considered structures. In the diamond case it is assumed that the lattice is un-defected, in that it contains no vacancies, interstitial atoms, heteroatoms or dislocation defects, all of which are common in naturally occurring and synthetic diamond and nanodiamonds. For a-C(:H) nanoparticles it was simply assumed that the structures are circumferentially discretised by butadiene units (--CH=CH--CH=CH--) but in reality their intricate amorphous structures will exhibit a much more complex mix of C$-$C, C$=$C and aromatic C$\simeq$C bonds almost certainly leading to less idealised and more irregular discretisation steps. Ideally, all of the adopted simplicities and structural irregularities should be considered in detail. Unfortunately doing so would open up the need to explore poorly constrained parameter spaces and, perhaps more importantly, would detract from the clearly na\"{i}ve but illustrative and useful simplicity of the current approach.

\begin{table}
\caption{The modelled bands in the DIB range ($\lambda \simeq 425-877$\,nm).}
\centering
\begin{tabular}{lccc}
      &         &        &       \\[-0.35cm]
\hline
\hline
      &         &        &        \\[-0.35cm]
  transition              &    wavelengths     &    radii   &         No. of bands         \\[0.05cm]
      $n =$         &     [ nm ]     &     [ nm ]   &                      \\[0.05cm]
\hline
                              &                                     &                                     &                            \\[-0.30cm]
\multicolumn{3}{}{l} nanodiamond - 21 of 43 radii $\rightarrow$ 24 bands \\ 
  $1 \rightarrow  2$  &   457.38     --    837.22 & 1.02      --     1.38  &          9          \\
  $1 \rightarrow  3$  &   448.83     --    849.50 & 1.65      --     2.27   &          6          \\
  $1 \rightarrow  4$  &   433.33     --    780.80 & 2.22      --     2.98   &           5          \\
  $1 \rightarrow  5$  &   488.00     --    823.02 & 2.98      --     3.87   &           4          \\[0.1cm]

\multicolumn{3}{}{l} a-C(:H) - 109 of 133 radii  $\rightarrow$ 137 bands \\ 
  $1 \rightarrow  2$  & 434.81     --    845.56 & 1.19      --     1.66   &           14          \\
  $1 \rightarrow  3$  & 433.14     --    866.73 & 1.94      --     2.74   &           32          \\
  $1 \rightarrow  4$  & 431.23     --    870.71 & 2.65      --     3.76   &          55          \\
  $1 \rightarrow  5$  & 429.23     --    613.79 & 3.34      --     3.99    &          36          \\
\hline
\hline
     &      &         &         \\[-0.25cm]
\end{tabular}
\tablefoot{
For particle radii from 0.4 to 4\,nm for both nanodiamond and a-C(:H) CNDs. Note that 14\% [26\%] of the nanodiamond [a-C(:H)] CNDs  generate two bands, and only two, in the DIB wavelength range.
}
\label{table_DIB_list}
\end{table}

None of the modelling results presented here take into account the effects of surface states, defects and heteroatoms and the transitions that they engender.  This aspect clearly plays an important role in the extensive laboratory measurements on CNDs \cite[e.g.,][]{Giordano_etal_2023,Kumar_etal_2022,Liu_etal_2020,Liu_2020,Ozyurt_etal_2023}, which indicate that the associated transitions lie at wavelengths $\lesssim 600 $\,nm, that is in the orange, yellow, green, blue, and violet colour ranges (see Fig. \ref{fig_CND_3}). The heteroatom states include O atoms in hydroxyl  $-$OH, ether $-$C$-$O$-$C$-$, and carbonyl $>$C=O bonds, as well as  N atoms in amino -NH$_2$ and pyrrolic and pyridine five- and six-fold rings (see Fig. \ref{fig_E_bands}). All this complexity is beyond the scope of the present study but might provide fertile ground for dedicated laboratory experiments. These states will yield bands at visible wavelengths in addition to those described here and attributed to QC effects. Thus, the band spectrum of astrophysical CNDs is likely be more varied and richer that that described here.

\section{Discussion}
\label{sect_discussion}

The conceptual model presented here leads to a relatively large number of bands that, despite their inherent regularity, actually lead to what appear to be somewhat irregular inter-band spacing when considering multiple particle sizes and transitions from the ground state. It is this aspect that perhaps most intriguingly resembles the DIB band distribution in wavelength space. 

The derived synthetic CND spectrum is clearly not as rich as the observed DIBs but then neither is this two-component (nanodiamond and simplified a-C(:H) CND) model. Nevertheless, it does lead to a hint of band grouping or clustering that is observed in the DIB wavelength positions. This is illustrated by the DIB density plot in Fig. \ref{fig_CND_3} (black line), for comparison the a-C(:H) CND band density plot is also shown (black dashed line). Further, the method does not predict the larger gaps that are observed in the DIB spectra. However, as pointed out in the preceding section, this model does not yet include a completely realistic approach to the size discretisation of a-C(:H) CNDs, nor does it include heteroatoms, and so probably underestimates the likely number of bands. Given that the adopted methodology is unable to predict the transition intensities and band strengths it is not yet possible to make detailed quantitative comparisons with the DIBs. 

An important aspect of this modelling approach is that it predicts DIB-like bands from distinct particles of the same generic species, and pairs of bands from single size particles, with the band positions being size dependent and, to a much lesser extent, structure, and composition dependent. Each given size of particle produces one band, or two at most, and hence band to band correlations would be expected to depend on environmental factors that act upon the dust size distribution.

Interestingly, the adopted approach can be tested for qualitative consistency with the four observed fullerene cation, C$^+_{60}$ DIBs. Assuming a fullerene cage radius of 0.6\,nm, it predicts $m_w^\star \simeq 0.3$ or 0.6, depending on whether the band origin is assumed to come from the $n=2 \rightarrow 1$ or $n=3 \rightarrow 1$ transitions, respectively. This is consistent with the general trends in the product $\{ m_w^\star \, d^2 \}$ that smaller particles have higher effective carrier masses as they approach the molecular domain, that is the carriers behave more like they exhibit the electron mass.

Effectively, any particle will show bands in the visible and this must also include any silicate, and SiC nanoparticles that might exist in sufficient abundance in the ISM. Given that SiC has a diamond-like structure, which in crystalline form will also show a step-wise size periodicity, it will  exhibit the kind of stepped band intervals as seen for the nanodiamond sizes of \cite{2020_Jones_nd_CHn_ratios}. The carbon and silicate grains in the ISM are predominantly amorphous and therefore unlikely to show the same size dependent regularity in their structures, as pointed out above, which implies that they might show other, perhaps more irregular size dependencies, in their bands. As with classical electronic excitation, the modelled transitions show a tendency to pile-up at short wavelengths ($\lambda < 600$\,nm).  If the pile-up bands were to overlap significantly this would be analogous to the increasing small particle extinction (absorption) at short wavelengths \cite[e.g.][]{2012A&A...542A..98J}. 
 
The nature of CT bands, arising from transitions between the aromatic moieties intrinsic to a-C(:H) materials and their possible relationship with DIBs is outlined in some speculative detail in Appendix \ref{sect_CT_DIBs}. Indeed the broad excitation dependent emission from CNDs can be explained by a mixture of small molecules with some charge transfer coupling \citep[e.g.,][]{scott_etal_2024}.

\section{Conclusions}
\label{sect_conclusions}

This CND approach, if it is at all close to reality, implies that it may difficult to get completely satisfactory DIB identifications for many of the bands through laboratory experiments because they are due to size discretisation (quantisation) effects rather than being good indicators of structure or  composition. On the other hand, this approach implies that, if CNDs make a significant contribution to the DIBs, then they must exhibit numerous and particularly stable discretised size states. 

If CNDs are responsible for a significant number of the DIBs and size, and aromatic domain shape, rather than exact chemical composition are the phenomenological determining factors, then we may have to come to terms with the possibility that the identification of the DIB carriers, far from being extremely difficult, may actually prove to be impossible for many, if not most, of them. However, and even though an exact chemical/structural identification might not be possible, with this approach we may be able to deduce some information about the size and/or the number of atoms associated with the DIB carriers. 

This modelling framework could provide an indication as to why the DIB problem has remained unresolved for so long: except in the unique case of the four long wavelength DIBs attributed to C$_{60}^+$. Unfortunately, and as satisfying as the C$_{60}^+$ DIB identifications may seem, the assignment of four bands to a single species does not constitute a widely applicable solution to the outstanding problem of the more than 550 other DIBs. 

As a consequence of this modelling it appears that if the valence and conduction band (VB, and CB) structures of CND nanoparticles do split into distinct levels at small sizes ($a < 10$\,nm) then the VB, andCB band approximations are no longer a good approximation for the determination of their optical properties. In this case we may have to completely re-think the way that we treat nanoparticles in astrophysics.

\begin{acknowledgements} 
The author wishes to thank the referee for valuable and insightful suggestions and acknowledges the time of Covid-19 confinement for providing a unique opportunity for thoughtful and fruitful reflection. 
\end{acknowledgements}

\bibliographystyle{aa} 
\bibliography{Ant_bibliography.bib}


\begin{appendix}

\section{Fluorescent emission from CNDs}
\label{sect_CD_emission}

The fluorescent emission covers the blue to red wavelengths, with the main emission bands occurring in the blue at $\sim 450$nm when excited in the $n-pi^\star$ absorption band ($\sim 350$nm). Long wavelength fluorescence ($>400$nm) also occurs due to radiative relaxation from excited to ground states due to O-containing groups on the surface. The origin of the fluorescent emission mechanism(s) is not always clear but surface state theories appear to be as widely accepted as QC. For particles with radii $\sim 1.0$ and 3.3nm the emission is in the blue and red, respectively, and  for radii of $\sim 3$ to 8\,nm a decrease in the optical gap from 3.0 to 2.2eV is associated with a redshift in the emission from the violet to the yellow. 

The carbon-core states control the emission of CNDs through radiative recombination of electrons and holes, arising from $\pi-\pi^\star$ transitions of $sp^2$ clusters assisted by a quantum confinement (QC) effect. QC is clearly significant in CNDs with extended conjugated $sp^2$ domains and fewer surface state functional groups. The fluorescence and electronic properties depend on electronic transitions across the band gap of $sp^2$ domains, affected by surface states, fluorophores and doping. Heteroatom dopants and defects within $sp^2$ domains can act as exciton capture centres and promote radiative relaxation from excited to ground states and thus affect the surface-related fluorescence. Oxygen and nitrogen heteroatoms lead to impurity levels in the band gap (see Fig. \ref{fig_E_bands}) which lead to transitions that can explain variations in the excitation and emission spectra. A redshift of the fluorescent emission with increasing particle size is consistent with QC but is also observed to correlate with N content, in particular graphitic N (N atoms replacing C atoms in the graphite plane) also leads to an emission redshift. 

Fluorescence from CNDs is generally single colour but multicolour emission, due to  fluorescent centres with different degrees of oxidation or due to N atoms on the surface, is also seen. The fluorescent emission from CNDs can show high quantum yields, $>70$\% and up to 99\%. Further, the fluorescent intensity monotonically decreases with increasing temperature.

\section{Discretisation of a-C(:H) nanoparticles}
\label{sect_aCH_CND}

The idealised circumference (and hence radius) discretisation of a-C(:H) nanoparticles is described here in detail.  

We make the reasonable hypothetical assumption that the particle periphery grows by C$_4$ units, which can incorporate into the structure by forming 6-fold rings (see Fig. \ref{fig_structures}). Adopting $j$ as the number of circumferential $-$C$=$C$-$C$=$C or, equivalently $=$C$-$C$=$C$-$C, bond units per a-C(:H) nanoparticle (or carbon nanodot, CND). The bond length associated with $j$ is $2 \times 0.202 = 0.404$\,nm (see Sect. \ref{sect_model}), the circumference discretisation length $L_{cd}$. All characteristic lengths are then by default in nm. We define a dimensioning factor, $f_j$, to account for the transformation in shape from 2D to 3D with increasing size, $f_j = [ 5 \, \sqrt{( j + 18 )} + 0.05 ]$, which yields $\sim 1, \frac{2}{3}$ and $\frac{1}{2}$ for $j = 10, 45$ and 100, respectively, corresponding to particle dimensions, ND($j$) of 2, 2.5 and 3 (see lower left box in Fig. \ref{fig_structures}), where the particle dimension ND($j$) $= 1+ (1/f_j)$. The  particle radius, $a_{\rm ND(j)}$, for any given diimension is 
\begin{equation}
a_{\rm ND}(j)  = \frac{ f_j \, L_{cd} \, j}{2 \pi} = \frac{ L_{cd} \, j}{2 \pi \, [{\rm ND(j)}-1]}.
\end{equation}
In order to calculate the number of peripheral C atoms in the particle we consider its 2D equivalent with radius $a_{\rm 2D}(j) = L_{cd} \, j / (2 \pi)$ and $N_{\rm circ,2D}(j) = 2 \times 2 \pi a_{\rm 2D}(j) / L_{cd}$ the number of C atoms in the circumference. 
We now define a dimension-adjusted circumference $N_{\rm circ,ND}(j) = N_{\rm circ,2D}(j) / {\rm ND(j)}$, which re-distributes the 2D circumference over the calculated particle dimensions ND (lower left box in Fig. \ref{fig_structures}). 
In order to derive the number of C atoms per nanoparticle, $N_{\rm C}$(j), the following expression is taken from Appendix A of \cite{2012A&A...542A..98J}, 
\begin{equation}
N_{\rm C} = 2500 \left( [a_{\rm ND}(j)]^{0.8} \right)^3 \left\{ \frac{ \rho_{\rm a-C(:H)} ( 1 - X_{\rm H})} {12 - 11 X_{\rm H}} \right\} \simeq 330 \, [a_{\rm ND}(j)]^{2.4} 
\end{equation}
where $a_{\rm ND}(j)$ is in nm, the density $\rho_{\rm a-C(:H)} = 1.6$\,g/cm$^3$ and assuming an H atom fraction $X_{\rm H} = 0.05$ typical of H-poor materials.  The number of C atoms in the particle interior is then $ N_{\rm int}(j) = N_{\rm C}(j) - N_{\rm circ,2D}(j)$. 
The $j$-dependent nanoparticle radius discretisation length $\delta a_d(j)$ is then, 
\[
\delta a_d(j) = \frac{ f_{j+1} \, L_{cd} \, (j+1)}{2 \pi} - \frac{ f_j \, L_{cd} \, j}{2 \pi}  
\]
\begin{equation}
\ \ \ \ \ \ \ \ \ \ \ = \frac{ L_{cd}}{2 \pi} \left\{ (  ( f_{j+1}  (j+1) - f_j  j ) ) \right\},  
\end{equation}
varying from $0.062-0.018$\,nm for $a_{\rm ND}(j)  = 0.4-4$\,nm.

\section{Charge transfer, and DIBs}
\label{sect_CT_DIBs}

This primarily concerns the action of UV photons ($E_{\rm h \nu} > 5$\,eV) with energies greater than the a-C(:H) band gap \citep[$E_{\rm g} \simeq 0.5-2.5$\,eV,][]{2012A&A...540A...1J,2012A&A...540A...2J,2012A&A...542A..98J}. These UV photons promote excitation-induced conductivity, that is, they induce electron transitions (electron-hole pair formation) between aromatic domains. If the UV photons have energies greater than the a-C(:H) band gap but smaller than that required to drive conductivity then the energy can be localised within single aromatic domains within low dimensional, cage-like, mixed aliphatic-olefinic-aromatic structures. For excitation-ionisation driven conduction the photon energies would need to exceed the ionisation potential for an aromatic (e.g., 6.97\,eV for tetracene, C$_{18}$H$_{12}$,  Fig. \ref{fig_structures}). Thus, if the photon energy can be localised, it is photons with energies $\sim 3-6$\,eV, which are not bond-breaking photon energies, that could induce exciton formation and lead to intra-grain aromatic domain charge charge transfer (CT). The result could be the following photo-reversible processes, where an arrow represents a UV photon absorption or emission event, in approximate order of decreasing probability (top to bottom and left to right), 
\[
[ \, {\varhexagon\varhexagon^0-\varhexagon\varhexagon^0} \, ]^{n\pm} \ \, \rightleftarrows \ [ \, {\varhexagon\varhexagon^+\hspace{-0.08cm}-\varhexagon\varhexagon^-} \, ]^{n\pm}  \  \rightleftarrows \ [ \, {\varhexagon\varhexagon^{2+}\hspace{-0.1cm}-\varhexagon\varhexagon^{2-}} \, ]^{n\pm}
\]
\[
[ \, {\varhexagon\varhexagon^+-\varhexagon\varhexagon^0} \, ]^{n\pm}  
\] 
\[
\hspace{0.8cm}\downarrow\uparrow
\] 
\[
[ \, {\varhexagon\varhexagon^0-\varhexagon\varhexagon^+} \, ]^{n\pm}  \ \rightleftarrows \ [ \, {\varhexagon\varhexagon^{-}-\varhexagon\varhexagon^{2+}} \, ]^{n\pm}
\] 
\vspace{0.01cm}
\[
[ \, {\varhexagon\varhexagon^--\varhexagon\varhexagon^0} \, ]^{n\pm} 
\] 
\[
\hspace{0.8cm}\downarrow\uparrow
\]
\[
[ \, {\varhexagon\varhexagon^0-\varhexagon\varhexagon^-} \, ]^{n\pm} \ \rightleftarrows \ [ \, {\varhexagon\varhexagon^{2-} \hspace{-0.15cm}-\varhexagon\varhexagon^+} \, ]^{n\pm}  \ \rightleftarrows \ [ \, {\varhexagon\varhexagon^{3-}\hspace{-0.15cm}-\varhexagon\varhexagon^{2+}} \, ]^{n\pm},  
\] 
where the overall grain charge of either sign $n\pm$ will be size dependent and larger for larger grains. The process of electron-hole excitation could in principle lead to the charge-driven fragmentation of nanograins because it would be independent of the sign and magnitude of the net charge. It would act on grains that are already highly charged and push localised regions within the particle over a Coulomb repulsive edge. However, this is probably only possible for highly charged grains, which means that CT excitons, while not being destructive, could play a role in dust physics. 

It appears that intra-molecular CT has not yet been considered as a source of DIBs. This, despite the fact that CT bands are characteristic features in many compounds and are often present in organic conjugated materials and also in complexes with unsaturated heterocyclic species. Structures that are present within the THEMIS a-C(:H) nanoparticles \citep{2012A&A...540A...1J,2012A&A...540A...2J,2012A&A...542A..98J}. 

If CT bands between N, and O carrying sub-structures or moities occur then, for the seven most likely N atom sites ($-$N$<$,\footnote{$-$N$<$ denotes a N atom with three trigonally arranged single bonds.} =N$-$ and their incorporation into pyrrolic 5-fold, and pyridine 6-fold rings, and $\equiv$N), and the six most likely O atom sites ($-$O$-$ and the 5-fold, and 6-fold ring equivalents, $-$OH, $>$C$=$O, and epoxide), the number of configuration combinations is 42, yielding $\gtrsim 42$ possible bands. If we now allow for the substitution of P for N, and S for O atoms, then the number of band transitions could exceed 160. Further, two related chemical species that could also play a role in CT transitions are the cyanate ion OCN$^-$ (cyanate salt IR band at $\sim 4.77\,\mu$m)\footnote{The isomeric fulminate CNO$^-$ group is highly unstable viz. explosive, and therefore highly reactive and expected to be of little relevance.} and the five membered ring isoxazole ($\leftarrow$C$=$C$-$N$=$C$-$O$\rightarrow$, where the arrows indicate a single, linking C$-$C bond).

Some of the key requirements for viable DIBs carriers would appear to be consistent with CT moities, in that they must be: widespread and consist of abundant elements (e.g. C, H, N, O, Si, Mg, Fe, S, and P), of a common hydro-carbon chemistry (with N, and O heteroatoms), stable polar (chromophore) structures or configurations, of limited spectral characteristics (i.e. few allowed transitions and/or excited states per carrier), on chemical/photolytic pathways that yield the same fragment sub-species (e.g. C$_2$, C$_3$, CN, CO, and HCO), and able to explain the abundant red, orange, and yellow DIBs. 

The carriers of the DIB-exhibiting grain (sub)structures or moities are most likely formed by a top-down process \cite[e.g.][]{2016RSOS....360223J} from the interstellar carbonaceous (nano)particles that may also carry the UV extinction bump. For example, in their study of 11 DIBs along 49 sightlines \cite{2005MNRAS.358..563M} found that the equivalent widths of some DIBs correlate positively with the extinction in the region of the 217.5\,nm bump, perhaps supporting this scenario. However, they also showed that correlations with colour excesses elsewhere in the UV  vary from one DIB to another, with both positive (579.7, 585.0 and 637.6\,nm) and negative (578.0 and 628.4\,nm) correlations possible. 
 
The DIB carriers must be (E)UV photon resistant, although some must be a little less stable against photolysis. They should also be stable against chemical attack by gas phase atoms and ions, such as C, C$^+$, O, N, S, and S$^+$ in the low density atomic ISM. Alternatively, reaction with these species may lead to stable states and configurations within the carriers. Given their likely hydrocarbonaceous nature they may be related to the extended red emission (ERE) and/or blue luminescence (BL) carriers \cite[e.g.][]{2014P&SS..100...26J}.

\section{Quantum wells}
\label{sect_wells}

A quantum well is a potential well with discrete energy values and occurs where particles, that were free in three dimensions, are confined to two. This occurs when the well dimension is comparable to the de Broglie wavelength of the carriers (electrons and holes), which then have discrete energy values. 

The `infinite' well model, where the walls are infinite, is the simplest model of this situation. A better description is often provided by the finite well model, which has wide applications to transistor junctions in the semiconductor industry. In this model the walls are modelled using a finite potential $V_0$, which is the difference in the CB energies of the adjacent semiconductors in the `sandwich'. The walls are finite and the electrons able to tunnel into the barrier region, with boundary conditions $\psi (0) = \psi (L) = 0$, where $L$ is the width of the well. 

In the ISM the particles are isolated, in a vacuum and the `infinite' well model may be more appropriate because the barrier, the grain surface, is likely impenetrable. The infinite 2D sheet well model could be applied to 2D particles such as Platt particles or planar polycyclic aromatic hydrocarbons where, for these disc-like structures, and along the direction of interest $z$, we could, simplistically take the width of the well, $L$, to be equivalent to the particle diameter ($L = 2a$) with the well width running from $z=0$ to $z=2a=L$ and $\psi (0) = \psi (2a) = \psi (L) = 0$. 

Another application could be to semiconducting a-C(:H) mantles, tens of atoms thick and of depth $d$, on particles of radius $R$, with the conditions that $d \lesssim 1$\,nm, and $d \ll R$, as in the THEMIS model. The 3D enclosed shell could be considered as a quantum well wrapped around on itself and the opposing barriers connected. In this case the region is barrier-less, closed, and approximated by a sphere of radius $R$  where the circumference ($2 \pi R$) is taken to be the well width in the `infinite' well model. The boundary conditions are cyclic, in all directions. With $\psi (0) = \psi (2 \pi R) = 0$ and setting $L = 2 \pi R$ we recover the above quantum well expressions $\psi (0) = \psi (2 \pi R) = \psi (L) = 0$, which runs from $z=0$ to $z = 2 a = 2 \pi R$. 

\begin{figure}
\centering
 \includegraphics[width=5.0cm]{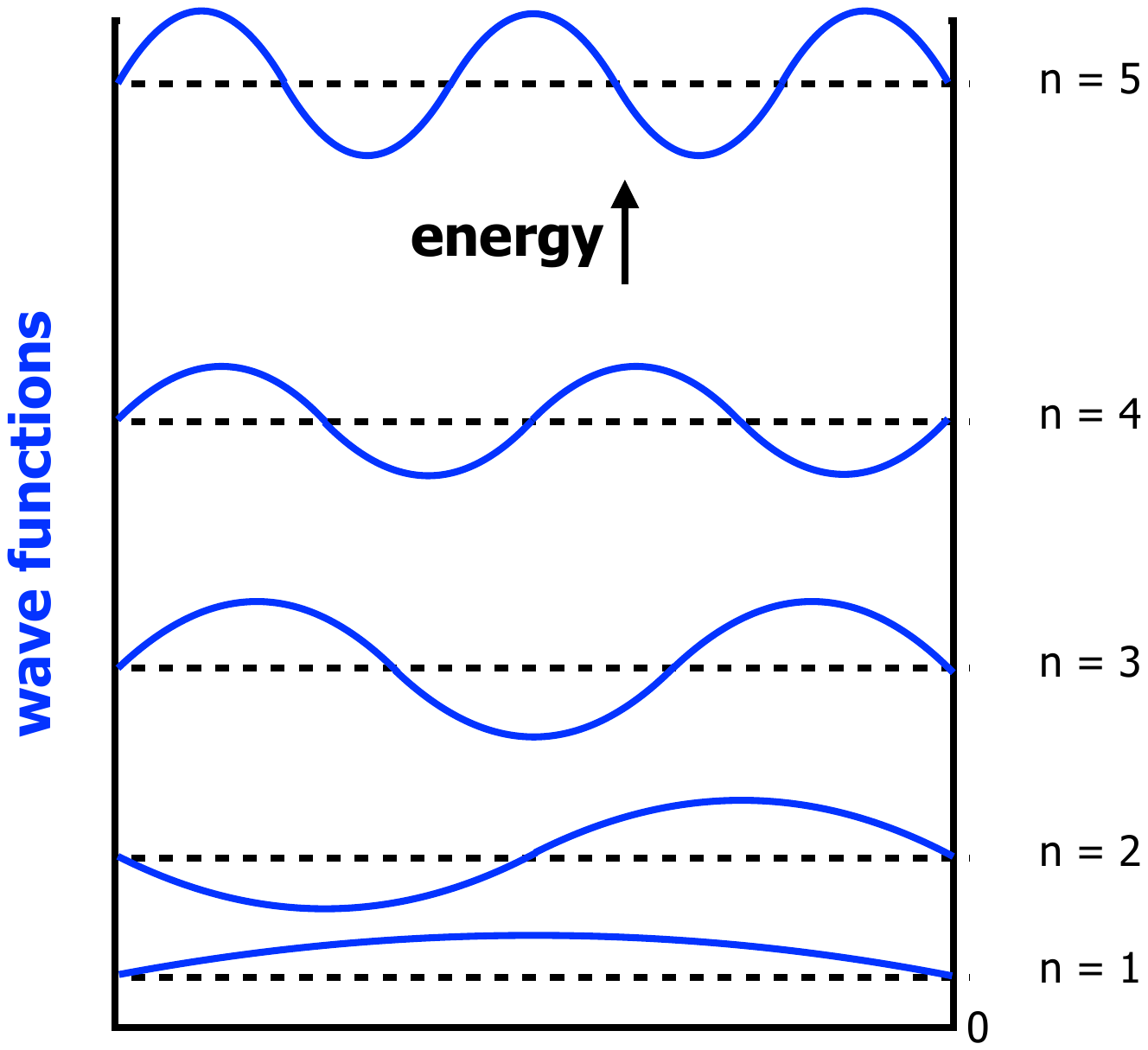}
 \caption{An `infinite' quantum well and its wave functions.}
 \label{fig_Q_well}
\end{figure}

For an `infinite' well model the ``particle in a box model'' can be considered using the Schr\"{o}dinger equation in one dimension for either the electron or the hole associated with an exciton 
\begin{equation}
\frac{-\hbar^2}{2\, m_w^\ast} \frac{d^2\phi_n}{dz^2} + V_0(z) \, \phi_n = E_n \, \phi_n,  
\end{equation}
where where $V_0(z)$ is the potential seen by the particle along $z$, $m_w^\ast$ is the effective particle mass, and $E_n$ and $\phi_n$ n are the eigenenergy and eigenfunction for the $n^{\rm th}$ solution to the equation. The barriers of the quantum well are infinitely high, the wavefunction is zero at these walls, and the solutions are    
\begin{equation}
E_n = \frac{\hbar^2}{2\, m_w^\ast} \left[ \frac{n \, \pi}{L} \right]^2 = \frac{\hbar^2\,k_n^2}{2\, m_w^\ast},  
\end{equation}
where $k_n = (n \, \pi ) / L$ is the wave vector associated with each state and $n$ ($= 1, 2, 3,$ \ldots) is the integer quantum number,  which characterises the wavefunctions as illustrated in Fig. \ref{fig_Q_well}. 

Nevertheless, quantum dot physics appears to be infinitely more applicable to the case of nm-sized, isolated ISM dust particles. 

\end{appendix}

\end{document}